\documentclass[twocolumn,showpacs,preprintnumbers,amsmath,amssymb]{revtex4}

\usepackage{graphicx}
\usepackage{dcolumn}
\usepackage{bm}

\begin{document}

\title{Nonlinear model of source of a elastic field.}

\author{Bohdan Lev}

\affiliation{Department of the Synergetic , Bogolyubov Institute
of the Theoretical Physics, NAS Ukraine, Metrologichna 14-b, Kyiv
03143, Ukraine.}

\date{\today}

\begin{abstract}

A general description of the long-range elastic interaction is
proposed. The far-field type of the interaction is determined by
the way of symmetry breaking of the distribution of the elastic
field produced by the topological defect as isolated inclusions.
Every topological defect can be present as the source of the
elastic field and can be described in the terms of this field. At
the short distance the source can be described as nonlinear
object which present charge of linear theory of elastic field at
the far distance. In this article the nonlinear models of source
of a elastic field was proposed.

\end{abstract}
\maketitle

In the general case we have the ground state of the continuum
which can always be described in terms of the elastic field. In
the ground state, this field has some definite value which does
not depend on the point of the elastic continuum. The different
fluctuation of elastic field which mean value is zero present the
ground state too. The elastic field can be characterized by
various geometrical objects, i.e., a scalar if this field
describes the phase transition in a condensed matter, a vector
potential in electrodynamics, a second-rank tensor in the general
relativity etc. To consider the elastic field that determines this
continuum, we have to describe possible deformations of the
distribution of this field. The basic concept is that in any
system with broken continuous symmetry exist states in which the
elastic variables describe distortions of spatial configurations
of the ground state. These distortions arise if the continuous
symmetry of the elastic-field distribution is broken in a local
area. The first way to break the continuous symmetry is to
introduce a foreign inclusion in the elastic matter. Then the
long-range interaction between inclusions is determined by the
symmetry of the deformation of the elastic field. This
deformations are produced by the inclusions and interaction can be
expressed in terms of the characteristics of this inclusion. In a
system with broken continuous symmetry also can exist defects in
the elastic-field distribution, i.e., the topological defects.
Each topological defect has some core region, where the
elastic-field distribution is strongly destroyed, and a far-field
region where the elastic variable slowly changes in space. Both
inclusions and topological defects are foreign to the elastic
field and can not be described in terms of this field.
Nevertheless they must have characteristics which influence on
the ground state of the elastic continuum. This characteristics
determine the value and character of the deformation. Then the
amount value of various elastic deformations can be associated
with the effective interaction between inclusions. In other
words, the presence of the field deformation immediately leads to
the interaction. Such interaction did not need the carrier.

Let us consider how the deformation of the elastic-field
distribution can produce interaction of additional inclusions or
topological singularities. We start with the description of a
scalar field with spatially uniform ground state. We have to
describe probable deformations of the distribution of a scalar
elastic field $\varphi(\overrightarrow{r})$. With an additional
inclusion being introduced in the elastic matter, we can write
the action of this system in the form
\begin{equation}
S=\frac{1}{8\pi}\int\left\{
\left(\overrightarrow{\nabla}\varphi(\overrightarrow{r})\right)^{2}-2\sum_{i}f_{i}\varphi(\overrightarrow{r_{i}})\right\}d\overrightarrow{r}
\end{equation}
where the first term describes the deformation energy of the
elastic field $\varphi$ and the second term is responsible for
the effect of this inclusion, located at the point
$\overrightarrow{r_{i}}$, on the elastic field. Note that the
inclusion must possess properties which influence on the elastic
continuum. The first way to describe the interaction of
inclusions is to obtain the change of the deformation energy
produced by this additional inclusions. In the Fourier
presentation $\varphi(\overrightarrow{k})=\frac{1}{(2\pi)}\int
d\overrightarrow{r}\varphi(\overrightarrow{r})exp(-i\overrightarrow{k}\overrightarrow{r})$
we can rewrite the action in the $\overrightarrow{k}$ space in the
form
\begin{equation}
S=\frac{1}{8\pi}\frac{1}{(2\pi)^{3}}\int d\overrightarrow{k}
\left\{\overrightarrow{k}^{2}\varphi^{2}(\overrightarrow{k})-
2\sum_{i}f_{i}\varphi(\overrightarrow{k})exp(i\overrightarrow{k}\overrightarrow{r_{i}})\right\}
\end{equation}
In order to find probable configurations of the field $\varphi$ we
have to solve Euler-Lagrange (EL) equations with minimum of
action with regard to the boundary conditions on the inclusion.
This equation is given by
\begin{equation}
\overrightarrow{k}^{2}\varphi(\overrightarrow{k})=4\pi
\sum_{i}f_{i}exp(-i\overrightarrow{k}\overrightarrow{r_{i}})
\end{equation}
which corresponds equation $\Delta \varphi=-4\pi\sum_{i}f_{i}$ in
the real space. The solution of this equation yields the field
distribution in the form
\begin{equation}
\varphi(\overrightarrow{k})=-4\pi\sum_{i}f_{i}
\frac{exp(-i\overrightarrow{k}\overrightarrow{r_{i}})}{\overrightarrow{k}^{2}}
\end{equation}
Taking this distribution of the field in action again, we can
calculate the elastic-field deformation energy produced by two
inclusions:
\begin{equation}
U_{ij}=\frac{4\pi}{2}\frac{1}{(2\pi)^{3}}\sum_{i,j}f_{i}f_{j}
\int d \overrightarrow{k} \frac {exp
-i\overrightarrow{k}(\overrightarrow{r_{i}}-\overrightarrow{r_{j}})}{\overrightarrow{k^{2}}}
\end{equation}
Having integrated over $\overrightarrow{k}$ we obtain the
interaction energy of two inclusions in the standard form
\begin{equation}
U_{ij}=\sum_{i,j}\frac{f_{i}f_{j}}{\overrightarrow{r_{i}}-\overrightarrow{r_{i}}}
\end{equation}
If we assume that the scalar field is the electrostatic potential
and $f$ is the charge, we obtain the energy of Coulumb-like
charge interaction through the deformation of the electric field.
In order to take into account the charge dispersion in the matter
we have to describe the distribution of this field in the area of
dispersed charges. In this case, in this local area the symmetry
of the elastic-field distribution is different and
$\varphi(\overrightarrow{r_{i}})\simeq
\varphi(\overrightarrow{r})+(\overrightarrow{\rho_{i}}
\nabla)\varphi(\overrightarrow{r})$, where
$\overrightarrow{\rho_{i}}$ is the distance to a single charge
and, having introduced the dipole moment
$d_{i}=\sum_{i}b_{i}\rho_{i}$, we can find the dipole-dipole
interaction in standard form. The symmetry of distribution of
elastic field which produced by two inclusions is different as
symmetry of deformation of elastic field each inclusion.
Two-particle system selects the solution which corresponds to the
position and properties of these inclusions.

Let us find this solution for a dynamical electromagnetic field.
The action for electrodynamics may be written in the standard
form, i.e.,
\begin{equation}
S=-\frac{1}{16\pi c}\int
\left\{F_{ij}F^{ij}+\frac{16\pi}{c}A_{i}j^{i}\right\}d \Omega
\end{equation}
where the Maxwell stress tensor $F_{ij}=\frac{\partial
A_{i}}{\partial x_{j}}-\frac{\partial A_{j}}{\partial x_{i}}$ is
determined by the vector-potential $\overrightarrow{A}$, and
$\overrightarrow{j}$ is the current of charges. From the minimum
of action, we obtain the field equations:
\begin{equation}
\frac{\partial F_{ij}}{\partial x_{j}}=-\frac{4\pi}{c}j_{i}
\end{equation}
For the gauge condition $\frac{\partial A_{i}}{\partial x_{}}=0$
in the four-dimensional Euclidean space, the equation became wave
equation for Fourier-transformed vector potential
$A_{i}(k,\omega)$ with right term $j_{i}$, whose solution can
write as $A_{i}(k,\omega)=\frac{4\pi c j_{i}exp i
(\overrightarrow{k}\overrightarrow{r}+\omega
t)}{c^{2}k^{2}-\omega^{2}}$. The substitution of this solution
into the expression for the action give the interaction energy of
different currents in the standard form, i.e.,
\begin{equation}
U_{i,j}=\int d^{4}q j_{i})G^{ij}(q)j_{j}(q)
\end{equation}
where $G(\mathbf{q})$ is the Green function. For example using
this equation one can describe the interaction of two charges
which move with the velocity $v$. In this case the charge changes
by the law $e\delta(\mathbf{r}-\mathbf{v}t)$. In accordance with
the result \cite{lan}, the Fourier component of the vector
potential can be written as
\begin{equation}
\varphi_{\mathbf{k}}=4\pi
e\frac{exp(-i(\mathbf{k}\cdot\mathbf{v})t)}{k^{2}-(\frac{\mathbf{k}\cdot\mathbf{v}}{c})^{2}}
\end{equation}
and
\begin{equation}
A_{\mathbf{k}}=\frac{4\pi e}{c}\frac{\mathbf{v}
exp(-i(\mathbf{k}\cdot\mathbf{v})t)}{k^{2}-(\frac{\mathbf{k}\cdot\mathbf{v}}{c})^{2}}
\end{equation}
Having substituted this vector potential in the expression for
the interaction energy we find that in the case $v=0$ we have the
previous result. In the other case when $j_{0}=e(t)\delta
(\overrightarrow{r})$ yields the interaction energy in the form:
\begin{equation}
U(\overrightarrow{r}\overrightarrow{r'},t,t')=\frac{e^{2}\delta
(c(t-t')-(|\overrightarrow{r}-\overrightarrow{r'}|))}{|\overrightarrow{r}-\overrightarrow{r'}|}
\end{equation}
that reproduces the standard resultant interaction of variable
charges.

Same result we can obtain if describe the gravitation field and
the appearance of the interaction of masses which produce the
change of the geometry of space. The action of the gravitation
field and distributed matter is given by the standard expression
\begin{equation}
S=\frac{c}{16\pi G}\int R \sqrt{-g}d\Omega+\frac{1}{2c}\int T
\sqrt{-g}d \Omega
\end{equation}
where $R$ is the curvature, $T$ is the compression of the
energy-momentum tensor  with the metric tensor $g_{\mu \nu}$,
$\Omega$ is the space-time volume, and $G$ gravitational
constant. Minimization of this action yields the Einstein equation
\begin{equation}
R_{\mu \nu}-\frac{1}{2}g_{\mu \nu}R=\frac{8\pi G}{c^{4}}T_{\mu
\nu}
\end{equation}
For a distributed matter, the energy-momentum tensor can be
written as $T_{0 0}=-mc^{2}$ and the field equation reduces to
$R_{0 0}=-\frac{4\pi G}{c^{4}}T_{0 0}$. In the linear
approximation we have $\Gamma_{0 0}\simeq -\frac{1}{2}g_{\mu
\mu}\frac{\partial g^{0 0}}{\partial
x_{\mu}}=\frac{1}{c^{2}}\frac{\partial \varphi}{\partial
x_{\mu}}$ that have as result \cite{lan} $R_{0
0}=-\frac{1}{c^{2}}\frac{\partial^{2} \varphi}{\partial
x^{2}_{\mu}}=- \frac{1}{c^{2}}\triangle \varphi$. Thus we obtain
the Poison equation $\triangle \varphi =4\pi G m$. We substitute
the solution of this equation in the expression for the action
and thus obtain the interaction energy of the distributed matter
in the form of the standard Newton law
\begin{equation}
U =-\frac{G
m_{i}m_{j}}{\overrightarrow{r}_{i}-\overrightarrow{r}_{j}}
\end{equation}
This interaction energy has attractive nature by virtue of the
specifics of gravitation filed which is described by the
second-rank tensor.

In Refs. \cite{inf}, \cite{inf1} the field equation in the empty
space was written as
\begin{equation}
R_{\mu \nu}-\frac{1}{2}g_{\mu \nu}R=0
\end{equation}
according to Einstein's statement that the geometry can not be
mixed with matter. Solution of this equation for a continuum
distribution of matter does not exist. However, there exists a
solution with a singular point in the distribution of matter. In
this presentation we can obtain the same equation for the
gravitational field. The field equation determines the law of
motion in terms of the integral of motion of the surface that
surrounds this singularity. If we start from action
\begin{equation}
S=\frac{c^{3}}{16\pi G}\int R \sqrt{-g}d\Omega
\end{equation}
for free gravitation field we can obtain the energy-impulse
tensor as variational parameter which describe peculiarity in
geometry \cite{eding}.

We illustrate this approach on presentation of charge in
electrodynamic. We start from the action for free electrodynamics
field in the standard form, i.e.,
\begin{equation}
S=-\frac{1}{16\pi c}\int F_{ij}F^{ij}d \Omega
\end{equation}
The variation of this action is follow \cite{lan}
\begin{equation}
\delta S=-\frac{1}{4\pi c}\int \left\{\frac{\partial
F_{ik}}{\partial x_{k}}\delta A_{i}+\frac{\partial}{\partial x_{k}
}(F_{ik}\delta A_{i})\right\}d \Omega
\end{equation}
At the standard approach the last term can be neglected because
it can be reduced to the surface integral which disappear in a
reason that disappear potential on the surface. But, if we have
peculiarity in elastic field distribution it is not correct. This
surface integral is not zero on the surface of defects and on the
surface of the area where exist the electrodynamic field. One can
obtain this non zero term for the defects. For $x_{k}=x_{0}=ct$
the general presentation take the form
\begin{equation}
-\frac{1}{4\pi c}\int \left\{\frac{\partial}{\partial x_{k}
}(F_{ik}\delta A_{i})\right\}d \Omega=-\frac{1}{4\pi c}\int
\left\{\mathbf{\nabla}\varphi \delta \mathbf{A}\right\}d V
\end{equation}
This term can be rewritten in the other form in spherical
coordinates
\begin{equation}
-\frac{1}{4\pi c}\int \left\{\mathbf{\nabla}\varphi \delta
\mathbf{A}\right\}r^{2}dr d \cos\theta d\phi=-\frac{1}{c}\int
\left\{\mathbf{\nabla}\varphi \delta \mathbf{A}\right\}r^{2}dr
\end{equation}
If we take in to account that the solution for defect
$\mathbf{\nabla}\varphi=\frac{k}{r^{2}}\mathbf{r}$, we can
present the last presentation in the form
\begin{equation}
-\frac{1}{c}\int \left\{\mathbf{\nabla}\varphi \delta
\mathbf{A}\right\}r^{2}dr=-\frac{1}{c}\int \left\{k\delta
\mathbf{A}\right\}d\mathbf{r}
\end{equation}
or
\begin{equation}
-\frac{1}{c}\int k\delta \mathbf{A}\frac{d\mathbf{r}}{d
ct}d(ct)=-\frac{1}{c}\int k \frac{u_{r}}{c}\delta
\mathbf{A}d(ct)=-\frac{1}{c}\int \left\{\mathbf{j}\delta
\mathbf{A}\right\}d(ct)
\end{equation}
For $x_{k}=r$ we can obtain the other stream of energy through
the surface
\begin{equation}
-\frac{1}{c}\int \left\{\mathbf{\nabla}\varphi \delta
\varphi\right\}r^{2}d ct=-\frac{1}{c}\int k \delta \varphi d ct
\end{equation}
Combining both obtaining part we can write in four dimensional the
additional part of the action for the electrodynamic field
\begin{equation}
-\frac{1}{c}\int \mathbf{j} \delta \mathbf{A}dV d ct
\end{equation}
After this we must solve equation for  elastic field
\begin{equation}
\frac{\partial F_{ij}}{\partial x_{j}}=j_{i}
\end{equation}

The charge in electrodynamics, as well as the energy-momentum in
the general relativity, is foreign with respect to the field and
cannot be described in terms of either potentials or the
geometry. For a system with broken continuous symmetry, we can
consider a class of defects in the distribution of elastic field
that are called the topological defects. A topological defect can
play the role similar to a particle which changes the elastic
field. In the general case, each topological defect has a core
region, where the elastic-field distribution is strongly
destroyed, and a far-field region where the elastic variable
slowly varies in space. The boundary conditions are then
determined by the conditions on the core of the defect. In this
approach we can find the interaction energy of topological
defects. We can start from action (1) without additional force
$f$. The equation of minimum action in this case is given by the
Euler-Lagrange equation
\begin{equation}
\triangle \varphi(\overrightarrow{r})=0
\end{equation}
This equation has many solutions. The first solution, $\varphi=0$
or $\varphi=const$, is trivial and describes the ground state.
The particular solution of this equation, compatible to the
existence of a topological singularity, may be written as
$\varphi_{n}=(mr^{n}+kr^{-(n+1)})Y_{n}(\theta,\phi)$, and in the
case $n=0$ we have a spherically symmetric solution $ \varphi
=\frac{k}{r}$ where $k$ may be interpreted as the magnitude of
the topological charge. This solution can describe the
singularity in the topological behavior of the scalar field. As a
result, this solution has infinite eigenvalue, however the
interaction energy of such topological charges is finite. A
superposition of two solutions for the scalar field,
$\varphi(\overrightarrow{r})=\varphi^{1}(\overrightarrow{r})+\varphi^{2}(\overrightarrow{r})$,
which produce two topological charges and determine the
interaction energy given by $U_{int}\equiv
E(\varphi^{1}(\overrightarrow{r})+\varphi^{2}(\overrightarrow{r}))-E(\varphi^{1}(\overrightarrow{r}))-E(\varphi^{2}(\overrightarrow{r}))$
yields
\begin{equation}
U_{i j}=\frac{2}{8\pi}\int^{\infty}_{0} (\nabla
\varphi^{1})(\nabla \varphi^{2})d
\overrightarrow{r}=\frac{k_{i}k_{j}}{d}
\end{equation}
where $d$ is the distance between the topological charges. This
formula implies that two topological defects with like signs in
the elastic scalar field repel according to the Coloumb law. In a
more rigorous approach \cite{lupe} the interaction energy of two
defects in the two-dimensional have the same law.

In order to understand that can present this topological defects
return to action of distribution of scalar elastic field Eq 1. We
would like to have the linear theory outside the topological
defects. But general theory can be nonlinearity. The nonlinearity
must play role only inside of the defect. For the defects
description we can not use the standard theory with
characteristic $f_{i}$ as function of local elastic field. In
this case we can present $f_{i}=a\varphi-\frac{1}{2}b\varphi^{3}$
when $r\leq R$, and where $R$ is size of core of defect. In this
case inside the defect the equation can be present in the form
\begin{equation}
\bigtriangleup \varphi-a\varphi+b\varphi^{3}=0
\end{equation}
The solution of this equation is
$\varphi=(\frac{2a}{b})sech(\sqrt{a}r)$ and it have noncorrect
asymptotical behaviour. The standard approach of nonlinearity is
not good to present particle as topological defect. In more
realistic situation we can take the characteristic of defect
$f_{i}$ as function of gradient of local elastic field. Outside
the defect we must have the previous result and can present the
characteristic of defect $f_{i}$ in the form
$f=f-\sum_{n}a_{n}(\nabla \varphi)^{n}$. This series must have
only odd $n$. This correspond to the symmetry when states
$\varphi$ and $-\varphi$ are equal. For this case the different
equation for the action minimization can be obtained
\begin{equation}
\nabla(\nabla\varphi)-\sum_{n}a_{n}\nabla \left\{(\nabla
\varphi)^{n-1}\right\}\varphi+\sum_{n}a_{n}(1-n)(\nabla
\varphi)^{n}+f=0
\end{equation}
In the case $n=1$ we have the simple equation $\triangle
\varphi=-f$. In the case $n=3$ we have the nonlinear equation
\begin{equation}
\nabla (\nabla \varphi)(1-6a_{3}\varphi \nabla
\varphi)-2a_{3}(\nabla \varphi)^{3}+f=0
\end{equation}
which have exactly solution $\varphi=k r^{\frac{1}{2}}$ with
$k=\frac{1}{5a_{3}}$ when the nonlinearity is greater than linear
part. It is correct for the distance $r\leq r_{0}\equiv
k^{2}f^{-\frac{2}{3}}$. Outside this distance we have previous
behavior of the solution of the nonlinear equation and standard
presentation of the behaviour of the potential. Other way to
describe defect can be realize if we take the characteristic of
defect $f_{i}$ in the form $f_{i}=a\theta(r-r_{0})\varphi$ where
$\theta$ -- theta function, which is 1 at $r<r_{0}$ and 0 at
$r\geq r_{0}$ where $r_{0}$ -- size of the core. The action of the
scalar elastic field take the form:
\begin{equation}
S=\frac{1}{8\pi}\int\left\{
\left(\overrightarrow{\nabla}\varphi(\overrightarrow{r})\right)^{2}+a\theta(r-r_{o})\varphi^{2}\right\}d\overrightarrow{r}
\end{equation}
The minimum of action realized on the equation
\begin{equation}
\Delta \varphi\left\{1+a\theta\varphi^{2}\right\}+a\theta(\nabla
\varphi )^{2}\varphi^{2}+2a\delta(r_{o})(\nabla \varphi
)^{2}\varphi^{2}=0
\end{equation}
In the area $r\geq r_{o}$ we have equation
\begin{equation}
\Delta \varphi+2a\delta(r_{o})(\nabla \varphi
)_{r-r_{0}}^{2}\varphi_{r=r_{0}}^{2}=0
\end{equation}
that is similar the Poison equation with charge
\begin{equation}
2a\delta(r_{o})(\nabla \varphi
)_{r-r_{0}}^{2}\varphi_{r=r_{0}}^{2}
\end{equation}
In the area inside the core we have the equation
\begin{equation}
\Delta \varphi\left\{1+a\theta\varphi^{2}\right\}+a\theta(\nabla
\varphi )^{2}\varphi^{2}=0
\end{equation}
which have the first integral $(\nabla
\varphi)^{2}\left\{1+a\varphi^{2}\right\}$. The solution of this
equation with such first integral can be written in the form
$a\varphi\sqrt{1+a^{2}\varphi^{2}}+ln
\left|a\varphi+\sqrt{1+a^{2}\varphi^{2}}\right|=2ac$ which have
the next asymptotic behaviour: if $a\varphi<1$ $\varphi\approx
cr$ and vise versa if $a\varphi>1$ $\varphi\approx
(\frac{2cr}{a})^{\frac{1}{2}}$. The charge of particle tend to
zero in the center of the core of defect. Outside the core we have
the normal behavior of the elastic field. Nonlinearity in terms
of the gradient of elastic field can solve problem of source of
the one.

In that way we can present the characteristic of defect or charge
as nonlinearity formation which outside can be consider as defect
in linear theory of elastic interaction. Every action determine
elastic deformation which produce the topological defect. They
can be only topological singularities in the distribution of the
elastic field. Interaction of these singularities is governed by
the deformation of the elastic field. The presentation of the
particle as topological defect has two preferences. First of all,
in this case we have the law of conservation of the topological
charge and this not dependent on the nature of elastic field.
Second, we can estimate the size of the particle as the core of
this defect. The structure of this core is a structure of
particle. Every core must have the energy $mc^{2}$, where $m$ is
mass of particle or core of the defect. In the process of arise
of two particles must appears two topological defect with
different signs. In this case arise the energy of interaction
between them. We can observe two particle only in the case when
the distance between them will be larger two sizes of core. In
the condition of equality of energy $2mc^{2}=\frac{f^{2}}{2R}$ we
obtain that the size of core of size of particle
$R=\frac{f^{2}}{4mc^{2}}$. If $f$ is electric charge we obtained
the size of electron, as elementary particle in the form
\begin{equation}
R=\frac{e^{2}}{4mc^{2}}=\frac{e^{2}\hbar}{4mc^{2}\hbar}=\frac{1}{4}\alpha
\lambda
\end{equation}
$\alpha=\frac{e^{2}}{\hbar c}$ is constant of electric
interaction and $\lambda=\frac{\hbar}{mc}$ is Compton length. The
size of electron is the size of core of topological defect in
electrostatic field and this size is smaller as Compton length in
$\sim 450$ time. In the case of gravitation field
\begin{equation}
R==\frac{1}{4}\frac{m^{2}}{m^{2}_{p}}\lambda
\end{equation}
where $m_{p}=(\frac{\hbar c}{G})$ is Plank mass. The elementary
particle in this presentation has not structure.

Thus we obtain a general approach that makes possible to explain
well-known interaction modes for different cases by considering
various types of local symmetry breaking. The local continuum
symmetry breaking present the topological defect in elastic
field. The interaction between two defect have same form as
interaction between inclusions. The defect in distribution of
elastic field produce nonlinearity of behaviour of the elastic
field. Nonlinearity of elastic field is motivation of appearance
of the charge in electrodynamic and location of mass in general
relativity.

\end{document}